\documentclass[twocolumn,showpacs,preprintnumbers,amsmath,amssymb,epsfig]{revtex4}

\usepackage{graphicx}
\usepackage{dcolumn}
\usepackage{bm}
\usepackage{epsfig}

\usepackage{float}
\usepackage{amsmath}
\usepackage{epsfig,floatflt}
\usepackage{subfigure}
\usepackage[usenames]{color}

\begin{document}


\title{Observational effects of the early episodically dominating dark energy}


\author{Chan-Gyung Park${}^{1}$, Jae-heon Lee${}^{2}$,
       Jai-chan Hwang${}^{2}$, and Hyerim Noh${}^{3}$}
\affiliation{
         ${}^{1}$Division of Science Education and Institute of Fusion
                 Science,  \\
                 Chonbuk National University,
                 Jeonju 561-756, Republic of Korea \\
         ${}^{2}$Department of Astronomy and Atmospheric Sciences, \\
                 Kyungpook National University, Daegu 702-701,
                 Republic of Korea \\
         ${}^{3}$Korea Astronomy and Space Science Institute,
                 Daejeon 305-348, Republic of Korea}
\date{\today}


\begin{abstract}
We investigate observational consequences of the early episodically dominating
dark energy on the evolution of cosmological structures.
For this aim, we introduce the minimally coupled scalar field dark energy
model with the Albrecht-Skordis potential which allows a sudden ephemeral
domination of dark energy component during the radiation or early matter era.
The conventional cosmological parameters in the presence of such an early 
dark energy are constrained with WMAP and Planck cosmic microwave
background radiation data including other external data sets.
It is shown that in the presence of such an early dark energy the estimated
cosmological parameters can deviate substantially from the currently known
$\Lambda\textrm{CDM}$-based parameters, with best-fit values differing
by several percents for WMAP and by a percent level for Planck data.
For the latter case, only a limited amount of dark energy with episodic nature
is allowed since the Planck data strongly favors the $\Lambda\textrm{CDM}$
model.
Compared with the conventional dark energy model, the early dark energy
dominating near radiation-matter equality or at the early matter era results
in the shorter cosmic age or the presence of tensor-type perturbation,
respectively. Our analysis demonstrates that the alternative cosmological
parameter estimation is allowed based on the same observations
even in Einstein's gravity. 
\end{abstract}

\maketitle

\section{Introduction}

\begin{figure*}
\mbox{\epsfig{file=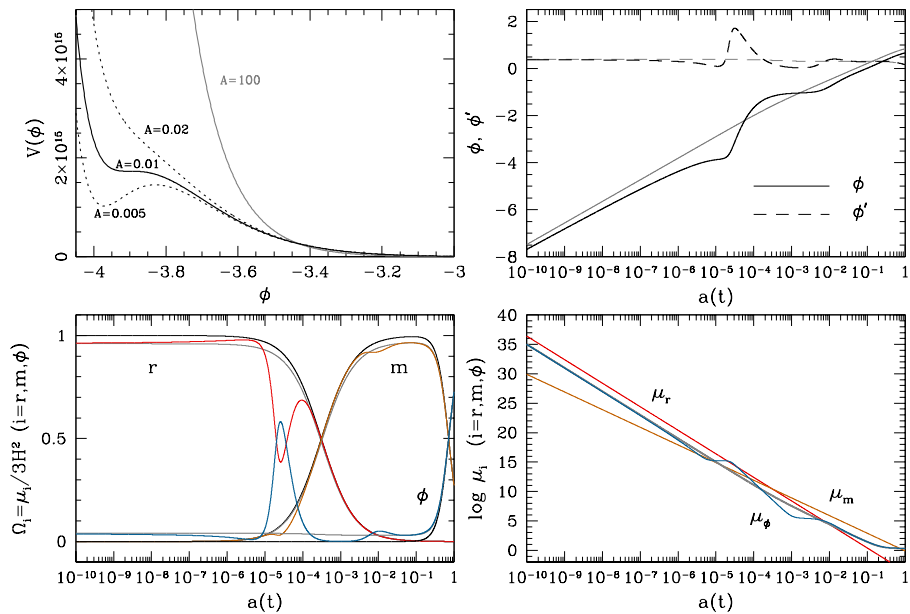,width=120mm,clip=}}
\caption{Top-left: A close-up picture of the scalar field potential
         [Eq.\ (\ref{eq:V}) in unit of $H_0^2$] in an EDE model with
         $\phi_0=-4$, $A=0.01$, and $\lambda=10$ (black solid curve).
         The cases for $A=0.005$ and $0.02$ are shown as dotted curves
         while that of SDE model ($A=100$, $\phi_0=-3$) as a gray curve
         with the potential amplitude suppressed by a factor of $300$
         for ease of display.
         Top-right: Evolution of $\phi$ and its time-derivative
         $\phi^{\prime}=d\phi/d\ln a$ in the EDE (black) and the SDE
         (gray curve) models as a function of the scale factor $a(t)$
         normalized to unity at present.
         Bottom: Evolution of background density parameters ($\Omega_i$)
         and energy densities ($\mu_i$) of radiation ($i=r$, red),
         matter ($m$, brown), and scalar field ($\phi$, blue curves)
         in the EDE model.
         The gray and black curves indicate the evolutions for the SDE and
         fiducial $\Lambda\textrm{CDM}$ models, respectively.}
\label{fig:bg_example}
\end{figure*}

The {\it precision cosmology} is often mentioned in present day cosmology
in the sense that the model parameters are constrained with a few percent
precision level \cite{Komatsu-etal-2011,Planck-2013}.
The future projects for cosmological observations also forecast the
sub-percent level parameter estimation in the optimistic situation
(e.g., see \cite{Euclid-2012} for Euclid project).
However, most of considerations for the observational precision rely on
the concordance cosmology based on the $\Lambda$CDM model with the
cosmological constant $\Lambda$ as the dark energy, the cold dark matter
(CDM) as the dominant dark matter, and the initial density and gravitational
wave power spectra provided by the inflation era in the early universe.
For the $\Lambda$, the effect of dark energy on the
evolution of universe at the early epoch is negligible or small.
Even in the dynamical dark energy models, attentions have been paid
on the late-time behavior of the dark energy that starts to dominate the
matter component only at very recent epoch, $z \lesssim 1$.

As soon as we introduce a dynamical dark energy, its prehistory could
become important \cite{Park-etal-2009}, and it is certainly allowed that
the dark energy may dominate at some earlier epoch without violating
observational constraints, but with differing cosmological parameter
estimations. In order to demonstrate such possibilities of alternative
cosmological parameter estimations in Einstein's gravity, here we would
like to investigate one such a model based on a known dark energy model
proposed in the literature as an example.
By introducing an early dark energy (EDE) which becomes ephemerally
dominant in the early cosmological era, we will show that
there exists room for alternative cosmological parameter estimation
based on the same cosmological observations.

A simple {\it fluid} model of EDE is known in the literature which
accommodates the scaling behavior of dark energy so that the dark energy
density parameter is constant before the onset of the late-time acceleration
\cite{Doran-2006}. Observational constraints on the fluid-based EDE model
were given by several groups using the cosmic microwave background radiation
(CMB), baryon acoustic oscillation (BAO), type Ia supernovae (SNIa),
Lyman-$\alpha$ forest, and Hubble constant data
\cite{Doran-2006,Xia-2009,Calabrese-2011,Joudaki-2011,Alam-2011,
Reichardt-etal-2012,Joudaki-2013}. The recent constraint on the EDE density parameter
from the Planck observation is $\Omega_e \lesssim 0.01$
(95\% confidence limit) \cite{Planck-2013}.
There are also studies on the transient acceleration models where the
universe might be decelerated in the near future
\cite{Transient-DE,Blais-2004}.

Here, we consider an early episodically (or ephemerally) dominating dark
energy model based on the scalar field which transiently becomes important
in the radiation- or early matter-dominated era before the onset of the
late-time acceleration.
Our aim is to investigate the observational effects and the consequent
cosmological parameter estimations based on the observationally
indistinguishable altered dark energy models.
Our field theory based model can accommodate the purely scaling behavior of
dark energy component and is free from the ambiguity seen in the fluid-based
EDE model. For the latter model, the behavior of the dark energy perturbation
variables strongly depends on the definition of sound speed and the method
of dealing with the dynamical property of the dark energy clustering.

This paper is organized as follows.
Section \ref{sec:EDE} describes the scalar-field-based EDE model adopted here.
We present the numerical calculations of background evolution of the EDE
model and provide the empirical method of obtaining the initial conditions
of the scalar field variables.
In Sec.\ \ref{sec:obs_eff}, we show how matter and CMB power spectra are
affected by the presence of EDE, and constrain the conventional cosmological
parameters for selected EDE models using the recent cosmological observations.
The summary and conclusion are given in Sec.\ \ref{sec:conclusions}.
Throughout this paper, we set $c \equiv 1 \equiv 8\pi G$.

\section{Early episodically dominating dark energy model}
\label{sec:EDE}

To mimic the episodic behavior of dark energy we consider a minimally coupled
scalar field with the potential
\begin{equation}
   V(\phi)=V_0 [(\phi-\phi_0)^2 + A] e^{-\lambda \phi}
          +V_1 e^{-\beta \phi}.
\label{eq:V}
\end{equation}
The first term is known as the Albrecht-Skordis potential with two interesting
points \cite{Albrecht-Skordis-model}:
(i) it allows the {\it near} scaling evolution where the energy density
of the scalar field scales with the dominant fluid density during the
evolution, and (ii) it drives the accelerated expansion of the universe.
It is known that both the permanent ($A\lambda^2 \lesssim 1$) and the
ephemeral ($A\lambda^2 \gtrsim 1$) accelerations are possible depending
on a choice of parameters $A$ and $\lambda$ \cite{Barrow-etal-2000}.
The previous studies on the transient nature of dark energy with the
Albrecht-Skordis potential were mainly focused on its recent transient
behaviors \cite{Blais-2004,Bean-etal-2001,Barnard-etal-2008}. On the other
hand, here we consider a different case in which the transient dark energy
domination occurs before the onset of the late-time acceleration that
will be driven by the second term of Eq.\ (\ref{eq:V})
(see \cite{Fedrow-2014} for a similar work). 
For simplicity, we set $\beta = 0$ throughout this work. Thus, the parameter
$V_1$ behaves as the cosmological constant $\Lambda$.

Figure \ref{fig:bg_example} shows an example of evolution of our EDE model,
where an extremely shallow local minimum was set by choosing $\phi_0=-4$,
$A=0.01$ for $\lambda=10$.  The value of $\lambda$ is related to the initial
dark energy density parameter as $\Omega_{\phi i}=4/\lambda^2$ during the
radiation-dominated era and has been chosen to satisfy the big bang
nucleosynthesis bound ($\Omega_{\phi i} < 0.045$)
\cite{Bean-etal-2001,HN-scaling-2001}. In this case the scalar field starts
to roll down the potential with scaling behavior, gets slower until it arrives
at the local minimum, and then suddenly kicks and rolls rapidly down the
potential out of the minimum. The rolling becomes slower as time goes by,
and the dark energy acquires the scaling nature again because the field rolls
down the potential which is effectively similar to the exponential one. We
consider a special case with a large value of $A=100$ ($\phi_0=-3$; gray
curves), which shows the effective scaling behavior without the episodic dark
energy domination. Hereafter, we refer to such a model as the scaling dark
energy (SDE) model.

\begin{figure*}
\mbox{\epsfig{file=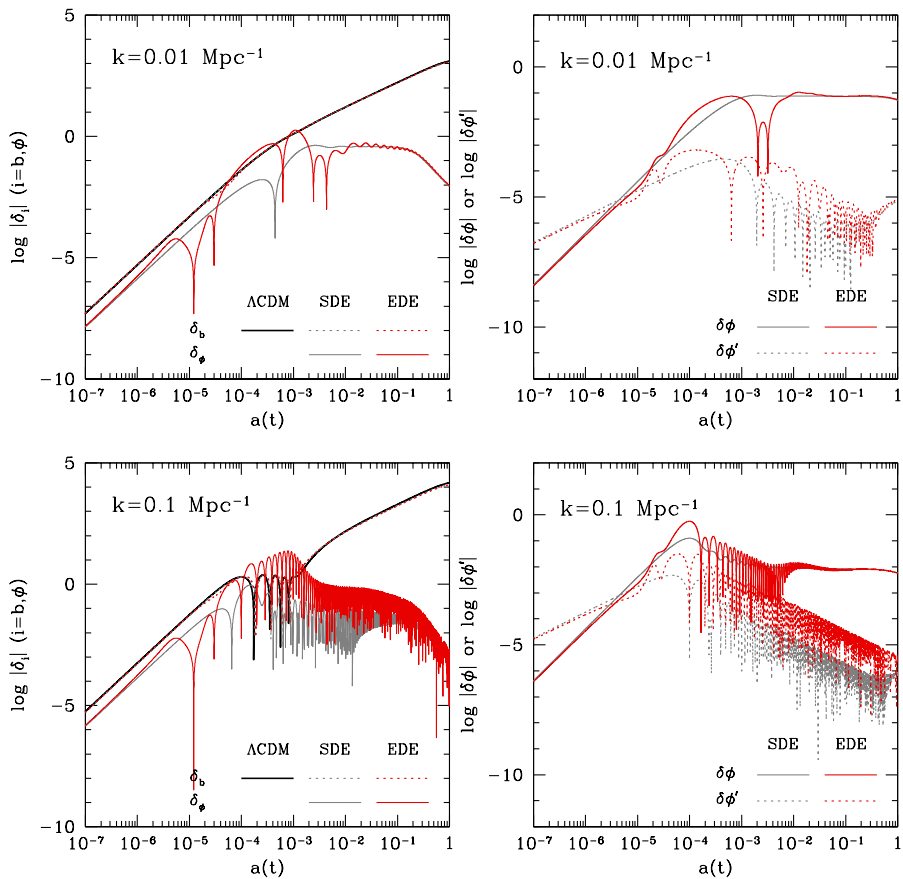,width=120mm,clip=}}
\caption{Evolution of density perturbations of baryon
         ($\delta_b$) and dark energy ($\delta_\phi$) components (left)
         and that of scalar field perturbation variables ($\delta\phi$
         and $\delta\phi^\prime=d\delta\phi/d\ln a$; right column)
         in the EDE model considered in Fig.\ \ref{fig:bg_example} (red curves),
         for comoving wavenumbers $k=0.01$ (top) and $0.1~\textrm{Mpc}^{-1}$ (bottom panels).
         As the temporal gauge condition, the CDM-comoving gauge (where the velocity
         of the cold dark matter vanishes) has been chosen.
         The gray and black curves indicate the evolutions for the SDE and
         fiducial $\Lambda\textrm{CDM}$ models, respectively.}
\label{fig:pert_example}
\end{figure*}

We evolve a system of multiple components for radiation, matter, and a
minimally coupled scalar field without direct mutual interactions among
the components. Our convention and the basic equations are summarized in
Refs.\ \cite{HN-scaling-2001,Hwang-2001-boltzmann}.
A set of initial conditions for the scalar field has been obtained with
an empirical method to give the scaling behavior of dark energy.
From the scaling evolution of the scalar field, we have energy density
and pressure as
\begin{equation}
   \mu_\phi \propto \mu_w,  \quad
     p_\phi = w\mu_\phi,
\label{eq:scalar_field_scaling}
\end{equation}
where $\mu_w$ is the energy density of the dominant fluid with
equation-of-state parameter $w$.
Combining the time-derivatives of $\mu_\phi$ and $p_\phi$ with
the equation of motion for scalar field gives a relation
\begin{equation}
   \phi^\prime =-3(1+w)\frac{V}{V_{,\phi}},
\label{eq:scaling_init_eq1}
\end{equation}
where a prime indicates a derivative with respect to $\ln a$.
Next, at the initial epoch $a_i=10^{-10}$ we choose a trial initial value
of $\phi_i$ within a sufficiently wide interval of $\phi < \phi_0$,
and obtain the initial condition for $\phi_i^\prime$ using the above relation.
From the scaling evolution of dark energy, we impose a condition that
the second derivative $\phi''$ should vanish at the initial stage
(numerically  $|\phi''|< 10^{-7}$; see the constant nature of $\phi^\prime$
for SDE model; Fig.\ \ref{fig:bg_example}, top-right panel).
By changing the value of $\phi_i$ using the bisection iteration technique
we obtain very accurate initial conditions for $\phi$ and $\phi^\prime$
within trials of about 30 times.
For each background evolution with these initial conditions,
the value of $V_1$ has been adjusted to satisfy the condition $H/H_0=1$
at the present epoch, where $H=\dot{a}/a$ is the Hubble parameter.
During setting these initial conditions, we simply set $V_0/H_0^2 \equiv 1$.
In the appendix, we present the initial conditions of background
scalar field variables ($\phi_i$, $\phi_i^\prime$) together with the potential
parameters of EDE (SDE) models considered in this paper.

\begin{figure*}
\mbox{\epsfig{file=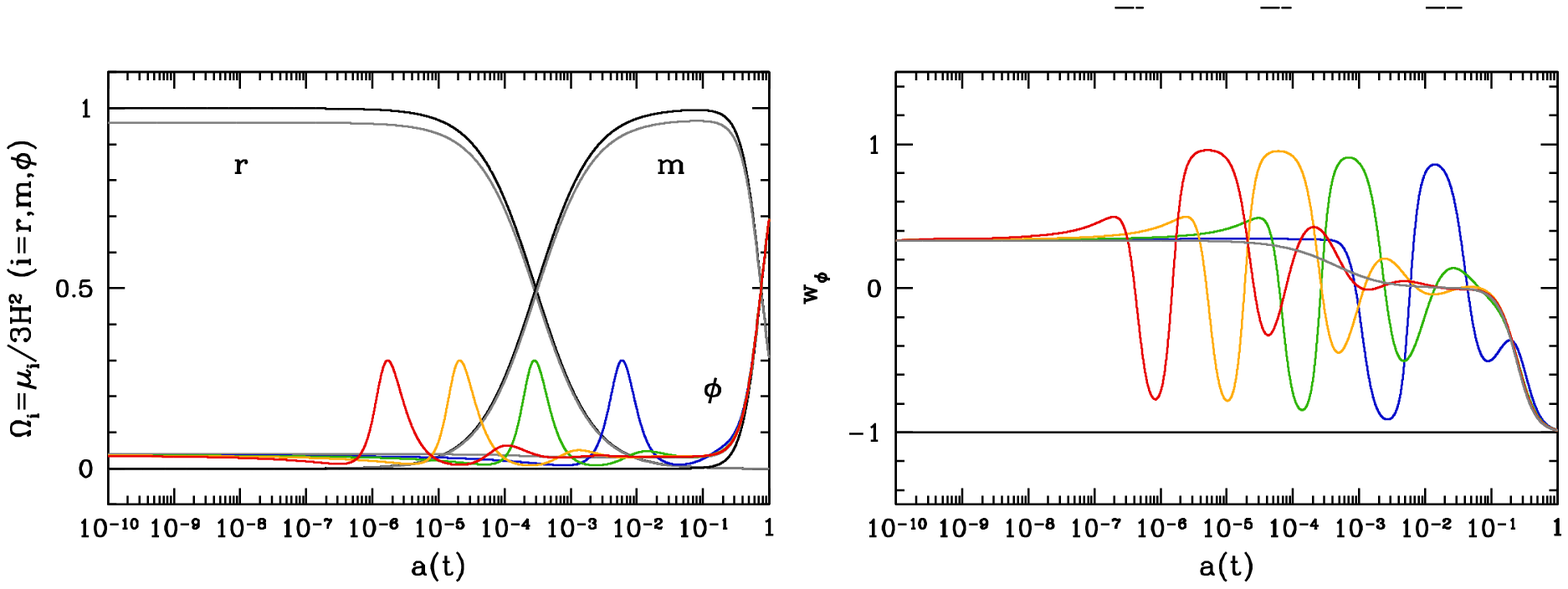,width=120mm,clip=}}
\mbox{\epsfig{file=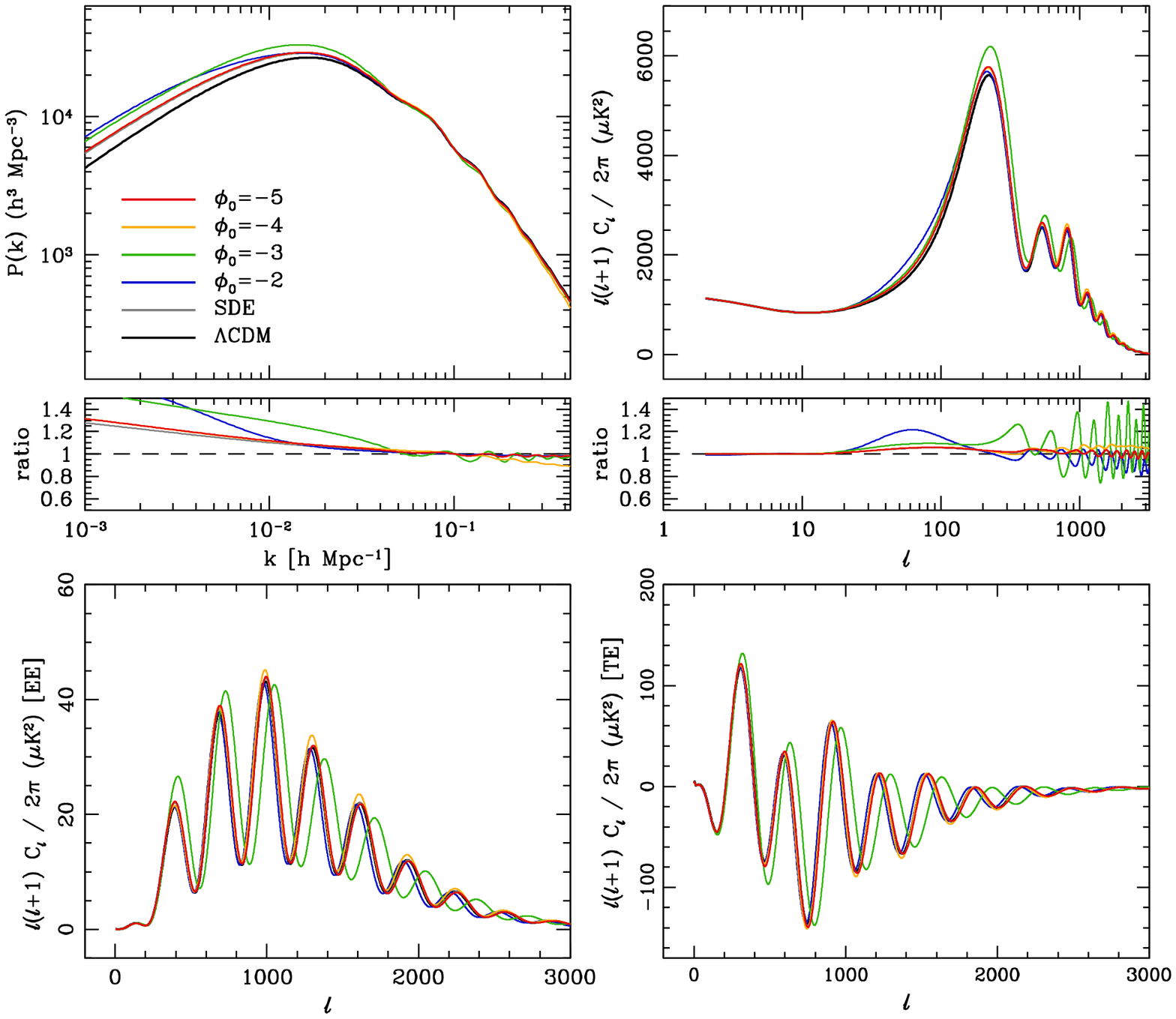,width=120mm,clip=}}
\caption{Evolution of density parameters ($\Omega_i$; $i=r$, $m$, $\phi$)
         and dark energy equation-of-state parameters ($w_\phi$) (top), and
         power spectra of baryonic matter density (middle-left), and of
         CMB temperature anisotropy (middle-right) and polarization
         (bottom panels) in our EDE models with $\phi_0=-5$ (red), $-4$
         (yellow), $-3$ (green), $-2$ (blue curves).
         For each value of $\phi_0$, $A$ has been adjusted to give
         $\Omega_{\phi}=0.3$ at the peak of early episodic domination
         of dark energy component. In all EDE models, we set $\lambda=10$.
         Grey curves represent the results of scaling dark energy (SDE) model,
         and black curves those of the fiducial $\Lambda\textrm{CDM}$ model.
         The matter power spectra are normalized to the $\Lambda\textrm{CDM}$
         model prediction at $k=0.1~h\textrm{Mpc}^{-1}$, while the CMB
         anisotropy power spectra at $\ell=10$.
         For matter and CMB temperature anisotropy power spectra,
         the ratio of EDE model power spectrum to $\Lambda\textrm{CDM}$
         prediction is also shown based on such a normalization.
         }
\label{fig:bgpert_examples}
\end{figure*}

For the scalar field perturbation variables, $\delta\phi$ and
$\delta\phi^{\prime}$, we have adopted the scaling initial conditions
for the pure exponential potential case \cite{HN-scaling-2001},
\begin{equation}
   \delta\phi=\frac{3(1-w)}{(7+9w)\lambda} \delta_w, \quad
   \delta\phi^\prime=2\delta\phi
\end{equation}
where $\delta_w \equiv \delta\mu_w / \mu_w \propto a^{1+3w}$
is the growing-mode solution of density perturbation variable for the
dominant $w$-fluid. These initial conditions are valid to use because
the background evolution of our EDE model shows very accurate scaling
evolution in the early era.
Although these are not the exact initial conditions, the actual evolution
of perturbation variables shows a scaling behavior quite well.

Figure \ref{fig:pert_example} shows evolution of density perturbations
of baryon ($\delta_b=\delta\mu_b/\mu_b$) and scalar-field dark energy
($\delta_\phi=\delta\mu_\phi/\mu_\phi$)
in the CDM-comoving gauge, together with that of scalar field perturbation
variables ($\delta\phi$, $\delta\phi^\prime$) at two different comoving
scales ($k=0.01$, $0.1~\textrm{Mpc}^{-1}$) for the EDE model ($A=0.01$,
$\phi_0=-4$, $\lambda=10$) in Fig.\ \ref{fig:bg_example}. 
Note that the evolution of baryon density perturbations (red dotted) slows down
during the episodic domination of dark energy (around $a \approx 3\times 10^{-5}$),
compared with the cases of $\Lambda\textrm{CDM}$ (black solid) and SDE models (gray
dotted curves; Fig.\ \ref{fig:pert_example} left panels), and that
the amplitude of dark energy perturbations becomes quite weaker 
during the recent epoch when the late-time acceleration is driven by the
cosmological constant.

\begin{figure*}
\mbox{\epsfig{file=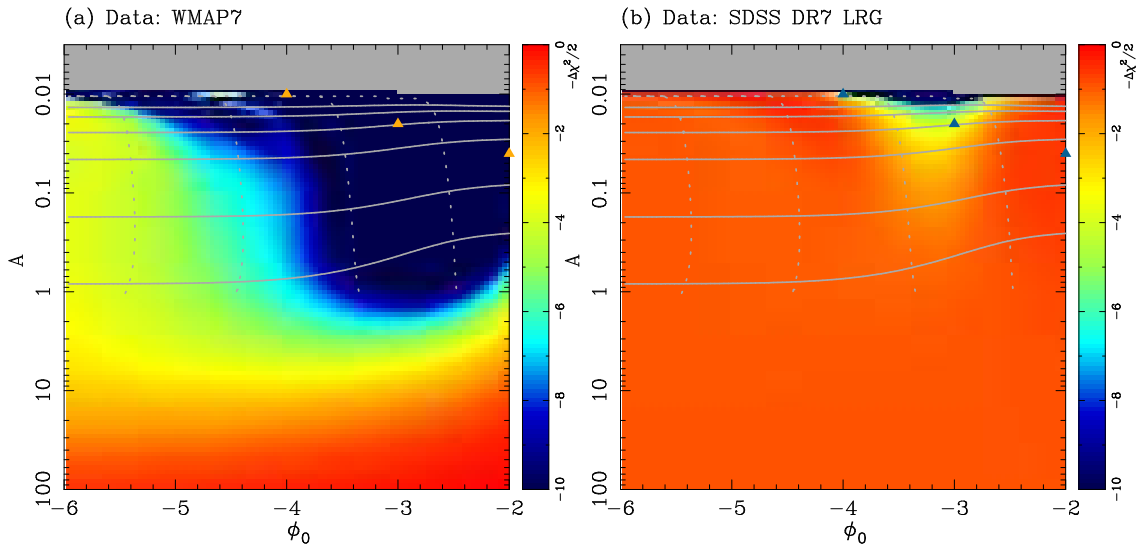,width=150mm,clip=}}
\caption{Probability distributions of potential parameters, $\phi_0$ and $A$,
         favored by ($a$) the WMAP 7-year data and ($b$) the SDSS DR7 LRG
         power spectrum.
         The gray region with $A \lesssim 0.01$ corresponds to the forbidden
         region where the permanent acceleration occurs.
         The relative logarithmic probability ($\Delta\ln P= -\Delta\chi^2/2$)
         decreases from red to dark blue.
         The dotted curves represent the epoch of EDE domination at maximum
         ($a=10^{-6}$, $10^{-5}$, $10^{-4}$, $10^{-3}$ from left to right)
         while the solid curves indicate the strength of dark energy
         at maximum ($\Omega_\phi=0.06$, $0.1$, $0.2$, $0.3$, $0.4$, $0.5$
         from bottom to top).
         Triangles indicate the potential parameters of three
         EDE models (EDE-1--3) chosen in the Markov chain Monte Carlo analysis
         (see Fig.\ \ref{fig:cont_EDE_nt_all} and Table \ref{tab:EDE_constraints_wmap}).
         }
\label{fig:prob_example}
\end{figure*}

\section{Observational constraints on early episodically dominating
dark energy model}
\label{sec:obs_eff}

\subsection{Observational signatures of scalar-field based EDE model}
\label{sec:model}

In order to see observational signatures of our scalar-field based EDE model,
we consider both the scalar- and tensor-type perturbations in a system of
multiple components for radiation, matter, and a minimally coupled scalar
field without direct mutual interactions among the components.
For this purpose, we modified the publicly available CAMB and CosmoMC packages
\cite{camb,cosmomc} by including the evolution of background and perturbation
of the scalar field quantities.

Figure \ref{fig:bgpert_examples} shows the evolution of background densities
and dark energy equation of state, baryonic matter density power spectrum,
and CMB temperature and polarization power spectra in the EDE model with
some chosen values of $\phi_0$. Here the strength of the ephemeral domination
has been fixed to $\Omega_{\phi}=0.3$ by adjusting the potential parameter $A$.
For smaller $\phi_0$ the epoch of the ephemeral domination occurs earlier.
We note that the dark energy component shows scaling behavior during the
radiation and matter dominated era except for the transient domination period
and that episodically dominating dark energy strongly affects the evolution
of perturbations. Figure \ref{fig:bgpert_examples} implies
that for the same episodic strength the EDE domination has weak observational
effects if it occurs before the radiation-matter equality (the cases of
$\phi_0=-5$ and $-4$).
However, the EDE domination near (or after) that epoch induces significant
deviations from $\Lambda\textrm{CDM}$ model prediction. 
For example, in the case of $\phi_0=-3$ which corresponds to EDE domination
near radiation-matter equality, we observe highly oscillatory features
at small angular scales in the CMB temperature anisotropy and the strong
deviation from $\Lambda\textrm{CDM}$ model at all angular scales in the
polarization power spectra.
Though the effects are much weaker, the same is true in the case of $\phi_0=-2$.
Compared with the CMB anisotropy, however, the matter power spectrum is less
sensitive to the presence of episodic domination of dark energy.

In the next two subsections, we explore the parameter constraints 
of the conventional cosmological parameters in the presence of episodic
domination of dark energy using the recent CMB data together with other
external data sets.

\subsection{Constraints from WMAP data}
\label{sec:wmap7}

\begin{figure*}
\mbox{\epsfig{file=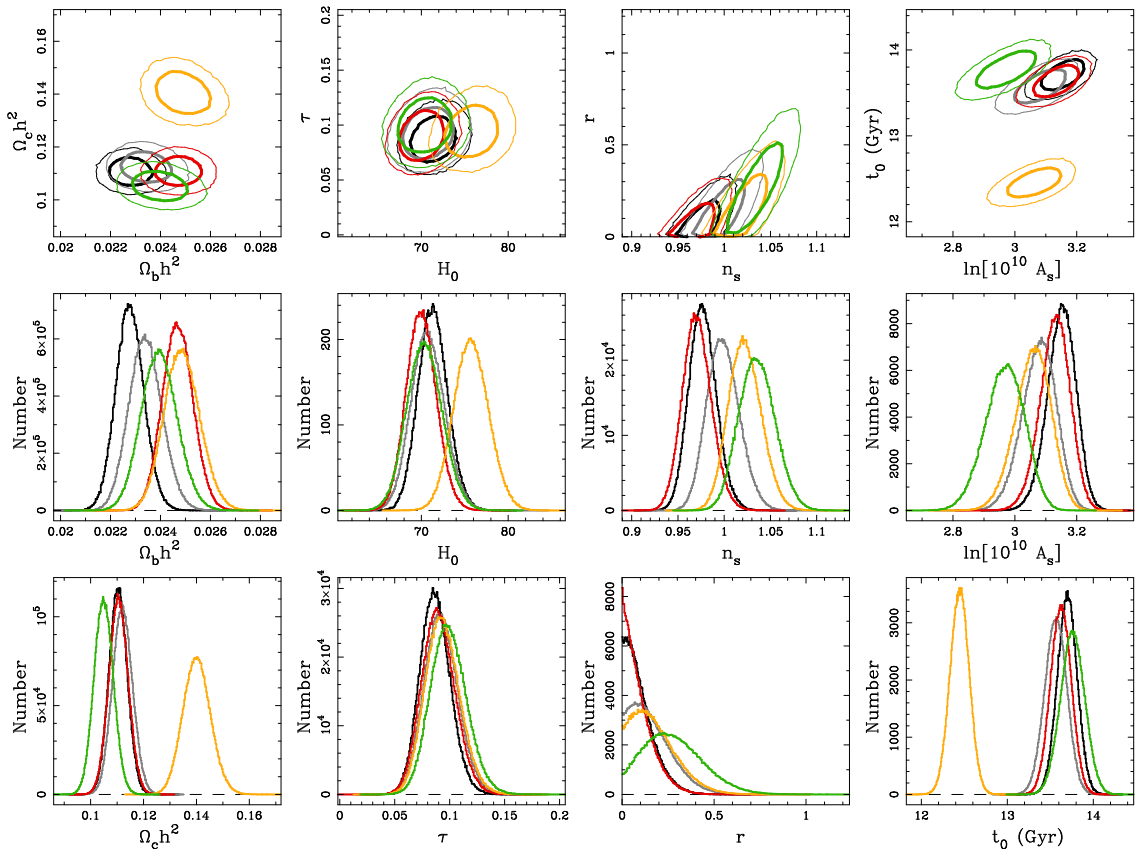,width=150mm,clip=}}
\caption{Top: Two-dimensional likelihood contours favored by WMAP7+LRG+$H_0$
         data sets for EDE-1  ($\phi_0=-4$, $A=0.01$; red),
         EDE-2 ($\phi_0=-3$, $A=0.02$; yellow), and EDE-3 ($\phi_0=-2$,
         $A=0.04$; green contours) models. For all models we set $\lambda=10$.
         The 68.3\% and 95.4\% confidence limits are indicated by the thick
         and thin solid curves, respectively.
         The results for the SDE ($\phi_0=-3$, $A=100$; gray) and
         $\Lambda\textrm{CDM}$ (black contours) models are shown for
         comparison.
         Middle and bottom: Marginalized one-dimensional likelihood
         distributions for each cosmological parameter.
         }
\label{fig:cont_EDE_nt_all}
\end{figure*}

First, we probe the overall ranges of potential parameters that are favored
by CMB and large-scale structure data.
Figure \ref{fig:prob_example} shows probability distributions of the scalar
field potential parameters obtained with the Wilkinson Microwave Anisotropy
Probe (WMAP) 7-year data \cite{Larson-etal-2011} and the Sloan Digital Sky
Survey Data Release 7 Luminous Red Galaxies (SDSS DR7 LRG) power spectrum
\cite{Reid-etal-2010}.
Here only the potential parameters, $\phi_0$ and $A$, have been probed
in a gridded space while other cosmological parameters are fixed with
the fiducial $\Lambda\textrm{CDM}$ best-fit values (see Table 14 of
\cite{Komatsu-etal-2011}). The relative logarithmic probability is defined as
$\Delta\ln P = -\Delta\chi^2/2 = -(\chi^2-\chi_\textrm{min}^2)/2$,
where $\chi_\textrm{min}^2$ is the minimum chi-square determined within the
area probed. It is shown that the epoch and strength of dark energy domination
are controlled by $\phi_0$ and $A$: the smaller $\phi_0$ ($A$) gives the
earlier (stronger) dark energy domination.
The presence of EDE has a major effect on the CMB anisotropy after the
radiation-dominated era ($a \gtrsim 10^{-4}$).
On the other hand, the matter density perturbation is generally less
sensitive to the EDE, being affected by only the strong EDE domination
near the radiation-matter equality ($a \approx 10^{-4}$--$10^{-3}$).
Therefore, as noted earlier, the parameter constraint are not much improved
by adding the galaxy power spectrum. In this example, we directly use the
galaxy power spectrum measured from the SDSS data.
During the model constraining, we have effectively excluded
the galaxy power spectrum information at scales $k>0.1~h \textrm{Mpc}^{-1}$
where the nonlinear clustering dominates.
The effect of bias parameter of galaxy clustering relative to the underlying
dark matter distribution is taken into account by marginalizing over
the overall amplitude of the galaxy power spectrum
(see Sec.\ 3 of \cite{Reid-etal-2010}).
We have not used the SNIa data because it is sensitive only to the
late-time acceleration of the universe and does not affect the behavior
of our EDE models with a limited range of $\phi_0 \le -2$.

In this example, the potential parameters with the larger value of $\phi_0$
and the smaller value of $A$ seem to be disfavored by the observational data.
However, we will see later that such potential parameters (denoted as triangles
in Fig.\ \ref{fig:prob_example} as an example) can be favored by the current
observations if different values of conventional cosmological parameters
are chosen.


\begin{table*}
\caption{Mean and standard deviation ($68.3$\% confidence limit) of
         cosmological parameters estimated from the marginalized
         one-dimensional likelihood distribution for best-fit
         $\Lambda\textrm{CDM}$, SDE, EDE models ($\lambda=10$)
         constrained with the recent observational data sets (WMAP7+LRG+$H_0$).
         For tensor-to-scalar ratio $r$, the upper limit or the peak-location
         in the likelihood is presented.
}
\begin{ruledtabular}
\begin{tabular}{lccccc}
  Parameter     & $\Lambda\textrm{CDM}$    &  SDE                 &  EDE-1             &  EDE-2         &  EDE-3        \\[1mm]
 $(\phi_0, A)$  &                          &  $(-3, 100)$         &  $(-4,0.01)$       &  $(-3,0.02)$   &  $(-2,0.04)$  \\[1mm]
\hline \\[-2mm]
  $100\Omega_{b}h^2$ & $2.281 \pm 0.056$   & $2.344 \pm 0.066$    & $2.473 \pm 0.061$  & $2.491 \pm0.071$ & $2.399\pm0.072$ \\[+1mm]
  $\Omega_{c}h^2$    & $0.1106 \pm 0.0035$ & $0.1123 \pm 0.0038$  & $0.1108 \pm 0.037$ & $0.1404\pm0.052$ & $0.1052\pm0.036$ \\[+1mm]
  $h$                & $0.714 \pm 0.017$   & $0.709 \pm 0.0019$   & $0.701 \pm 0.017$  & $0.758\pm0.020$  & $0.705\pm0.020$ \\[+1mm]
  $\tau$             & $0.088 \pm 0.014$   & $0.093 \pm 0.015$    & $0.091 \pm 0.015$  & $0.095\pm0.016$  & $0.100\pm0.016$ \\[+1mm]
  $n_s$              & $0.978 \pm 0.015$   & $1.000 \pm 0.017$    & $0.972 \pm 0.016$  & $1.024\pm0.018$  & $1.037\pm0.019$ \\[+1mm]
  $r$                & $< 0.128$           & $< 0.199$            & $< 0.109$          & $<0.225$         & $0.220_{-0.132}^{+0.179}$ \\[+1mm]
  $\ln[10^{10}A_s]$  & $3.153 \pm 0.046$   & $3.080 \pm 0.055$    & $3.132 \pm 0.049$  & $3.061\pm0.058$  & $2.970\pm0.063$ \\[+1mm]
  $t_0$ (Gyr)        & $13.70 \pm 0.12$    & $13.57 \pm 0.13$    & $13.63 \pm 0.12$    & $12.45\pm0.11$   & $13.77\pm0.14$ \\[+1mm]
\end{tabular}
\end{ruledtabular}
\label{tab:EDE_constraints_wmap}
\end{table*}

Next, we probe the conventional cosmological parameters in the presence of
EDE component and compare the results with those in the $\Lambda\textrm{CDM}$
model.
To obtain the probability (likelihood) distributions for those parameters,
we apply the Markov chain Monte Carlo (MCMC) method and randomly explore
the parameter space that is favored by the recent astronomical observations.
The MCMC method needs to make decisions whether it accepts or rejects a
randomly chosen chain element via the probability function
$P(\mbox{\boldmath $\theta$}|\mathbf{D}) \propto \exp(-\chi^2/2)$,
where $\mbox{\boldmath $\theta$}$ denotes a vector containing free model
parameters and $\mathbf{D}$ the data used, $\chi^2$ the sum of individual
chi-squares for CMB, large-scale structure data, Hubble constant, and so on.
We use the modified version of CAMB/CosmoMC software that includes 
the evolution of minimally coupled scalar field to obtain the likelihood
distribution of free parameters.
A simple diagnostic has been used to test the convergence of MCMC chains
(see Appendix B of Ref.\ \cite{Abrahamse-etal-2008}).

Depending on the epoch and strength of EDE domination, we consider three
spatially flat models where the dark energy dominates at radiation era
(EDE-1), near radiation-matter equality (EDE-2), and at early matter era
(EDE-3) by fixing $(\phi_0,A)=(-4,0.01)$, $(-3,0.02)$, and $(-2,0.04)$,
respectively, and $\lambda=10$. In each model, the value of $A$ has been
chosen so that we can obtain the chi-square between model and data that is
similar to that of $\Lambda\textrm{CDM}$ best-fit model.
Thus, the free parameters are $\Omega_{b}h^2$, $\Omega_{c}h^2$, $h$,
$\tau$, $n_s$, $r$, and $\ln[10^{10} A_s]$, where $\Omega_{b}$ ($\Omega_{c}$)
is the baryon (CDM) density parameter at the current epoch, $h$ is the
normalized Hubble constant with
$H_0=100 h~\textrm{km}~\textrm{s}^{-1}~\textrm{Mpc}^{-1}$,
$\tau$ is the reionization optical depth, $n_s$ is the spectral index
of the primordial scalar-type perturbation, $r$ is the ratio of
tensor- to scalar-type perturbations, and
$A_s$ is related to the amplitude of the primordial curvature perturbations
by $A_s=k^3 P_{\mathcal{R}}(k)/(2\pi^2)$ at $k_0=0.002~\textrm{Mpc}^{-1}$.
The running spectral index is not considered
(see Ref.\ \cite{Komatsu-etal-2011} for detailed descriptions of the
parameters)
\footnote{We choose the flat priors for the free parameters 
as $\Omega_b h^2 = [0.005, 0.100]$, $\Omega_c h^2=[0.01,0.99]$,
$h=[0.4,1.0]$, $\tau=[0.01,0.80]$, $n_s=[0.5,1.5]$, $r=[0,2]$,
$\ln (10^{10} A_s) = [2.7,4.0]$.}.
In order to constrain the model parameters we use the CMB
temperature and polarization power spectra measured from the WMAP 7-year data
\cite{Larson-etal-2011}, the galaxy power spectrum measured from the
SDSS DR7 LRG sample \cite{Reid-etal-2010},
and the recent measurement of the Hubble constant from the Hubble Space
Telescope ($H_0=74.2\pm 3.6~\textrm{km}~\textrm{s}^{-1}~\textrm{Mpc}^{-1}$;
\cite{Riess-etal-2009}).
From here on we denote the combined data sets as WMAP7+LRG+$H_0$.

The results of parameter constraints obtained with the MCMC method are
presented in Fig.\ \ref{fig:cont_EDE_nt_all}, which shows two-dimensional
likelihood contours and marginalized one-dimensional likelihood distributions
of cosmological parameters favored by WMAP7+LRG+$H_0$ data sets for three EDE,
SDE, $\Lambda\textrm{CDM}$ models.
Table \ref{tab:EDE_constraints_wmap} lists mean and $68.3$\% confidence limit
of cosmological parameters estimated from the marginalized one-dimensional
likelihood distributions. For tensor-to-scalar ratio $r$, however, we present
the location of the peak with $68.3$\% (upper) limit in the likelihood
distribution.

\begin{figure*}
\mbox{\epsfig{file=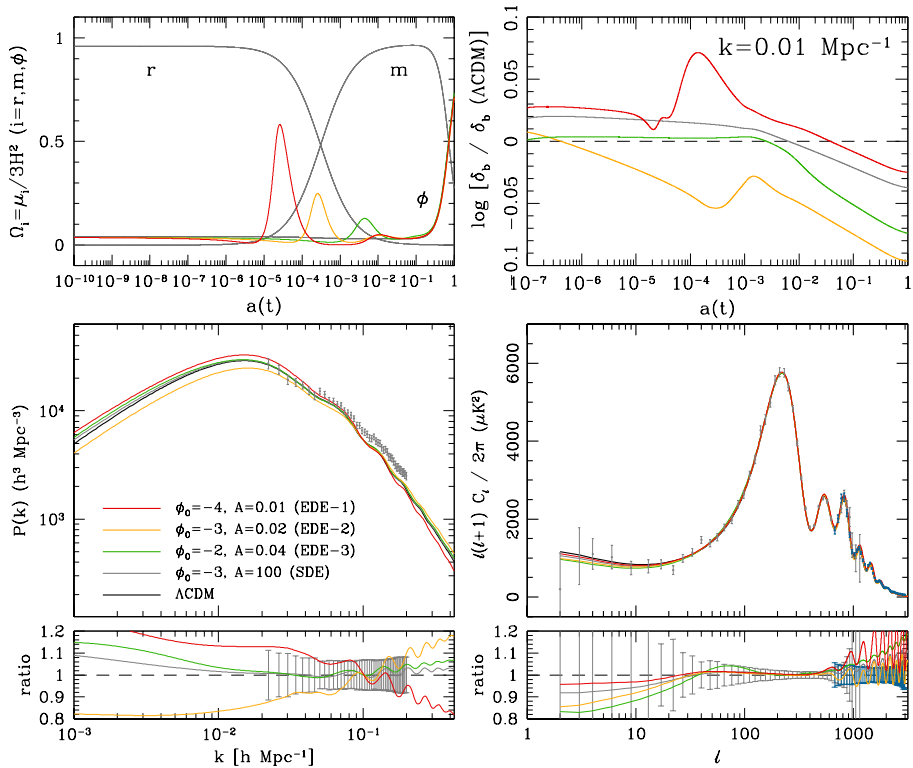,width=120mm,clip=}}
\caption{Evolution of background density parameters (top-left) and baryon
         density perturbations ($\delta_b \equiv \delta\mu_b /\mu_b$
         in the CDM-comoving gauge) at $k=0.01~\textrm{Mpc}^{-1}$
         relative to the $\Lambda\textrm{CDM}$ model (top-right),
         and baryonic matter and CMB anisotropy power spectra
         (sum of contributions from scalar- and tensor-type perturbations;
         bottom panels) for three best-fit EDE, $\Lambda\textrm{CDM}$,
         and SDE models, with the same color codes as in
         Fig.\ \ref{fig:cont_EDE_nt_all}.
         For matter and CMB power spectra, recent measurements from SDSS DR7
         LRG \cite{Reid-etal-2010} and WMAP 7-year \cite{Komatsu-etal-2011}
         data have been added (gray dots with error bars).
         In the ratio panels, we present the power ratio relative to 
         the $\Lambda\textrm{CDM}$ prediction together with the fractional
         error bars of the observational data including SPT data
         (at $\ell \gtrsim 600$; blue bars) \cite{SPT}.
         }
\label{fig:bgpert_nt_bestfit}
\end{figure*}

The parameter constraints for EDE-1 model are consistent with those for
$\Lambda\textrm{CDM}$ and SDE models, which implies that for our chosen
strength of the episode the EDE domination
at the early radiation era does not much affect the overall evolution of
density perturbations. On the other hand, we notice significant deviations
in the parameter constraints in EDE-2 and EDE-3 results.
First, in the case of EDE-2, some cosmological parameters at the best-fit
position significantly deviate from the $\Lambda\textrm{CDM}$ likelihood
distribution by over $3\sigma$:
compare the best-fit EDE-2 model parameters ($100\Omega_b h^2=2.491\pm 0.071$,
$\Omega_c h^2=0.1404\pm0.0052$, $n_s=1.024\pm0.018$)
with the $\Lambda\textrm{CDM}$ ones ($100\Omega_b h^2=2.281\pm 0.056$,
$\Omega_c h^2=0.1106\pm0.0035$, $n_s=0.978\pm0.015$).
Interestingly, the EDE-2 model fits data better than $\Lambda\textrm{CDM}$
model, that is, the minimum value of chi-square at the best-fit position 
in the EDE-2 model ($\chi_{\textrm{min}}^2/2=3748.6$) is smaller than the
$\Lambda\textrm{CDM}$ value ($\chi_{\textrm{min}}^2/2=3750.1$), and the
derived cosmic age ($t_0 = 12.45\pm 0.11~\textrm{Gyr}$) is quite smaller
than that of $\Lambda\textrm{CDM}$ model ($t_0=13.70\pm 0.12~\textrm{Gyr}$). 
Secondly, the EDE-3 model prefers the larger spectral index and
the positive tensor-to-scalar ratio ($n_s=1.037\pm 0.019$, $r=0.22\pm 0.16$),
implying that the presence of early episodic dark energy at early matter era
demands the existence of strong tensor-type perturbations unlike the best-fit
$\Lambda\textrm{CDM}$ model with $r$ consistent with zero.
In some sense, the large tensor-to-scalar ratio $r=0.20_{-0.05}^{+0.07}$
from the recent $B$-mode polarization measurement of the BICEP2 experiment
\cite{BICEP2} can be mimicked by the transient behavior of dark energy,
although such a large value of $r$ is not allowed in the EDE model
constrained by the Planck data (see next subsection).

Figure \ref{fig:bgpert_nt_bestfit} shows the evolution of background
and perturbation quantities, and the matter and CMB power spectra of
the three best-fit EDE models.
It should be noted that the CMB temperature power spectrum is the sum
of contributions from the scalar- and tensor-type perturbations.
The evolution of baryon density perturbations ($\delta_b=\delta\mu_b / \mu_b$
in the CDM-comoving gauge) relative to the $\Lambda\textrm{CDM}$ model
at comoving wavenumber $k=0.01~\textrm{Mpc}^{-1}$ shows that 
the growth of perturbation is affected by the episodic domination of dark
energy. Here the initial amplitudes of baryon density perturbation are
different among models because of the differently chosen best-fit initial
power spectrum amplitude ($A_s$).
The consequent EDE model power spectra, which are observationally
indistinguishable with $\Lambda\textrm{CDM}$ ones, suggests that 
the different parameter constraints with significant statistical deviations
from the $\Lambda\textrm{CDM}$ model can be obtained by introducing
the alternative dark energy model (here with the early episodically
dominating dark energy) even based on the same observational data.

The WMAP 9-year data \cite{WMAP9} will provide parameter constraints
a bit tighter than our WMAP 7-year result because the noise level of WMAP
9-year measurement decreases by a factor of $\sqrt{9/7}=1.13$ compared with
WMAP 7-year one without dramatic changes in systematic effects.
We expect that even tighter constraints on model parameters
are obtained if our EDE model is confronted to recent CMB data with higher
precision at small angular scales, e.g., from the South Pole Telescope (SPT)
\cite{SPT} (compare EDE model's deviations relative to $\Lambda\textrm{CDM}$ model
with the size of SPT error bars in Fig.\ \ref{fig:bgpert_nt_bestfit} bottom-right panel)
and Planck (next subsection).

\subsection{Constraints from Planck data}
\label{sec:planck}

\begin{figure*}
\mbox{\epsfig{file=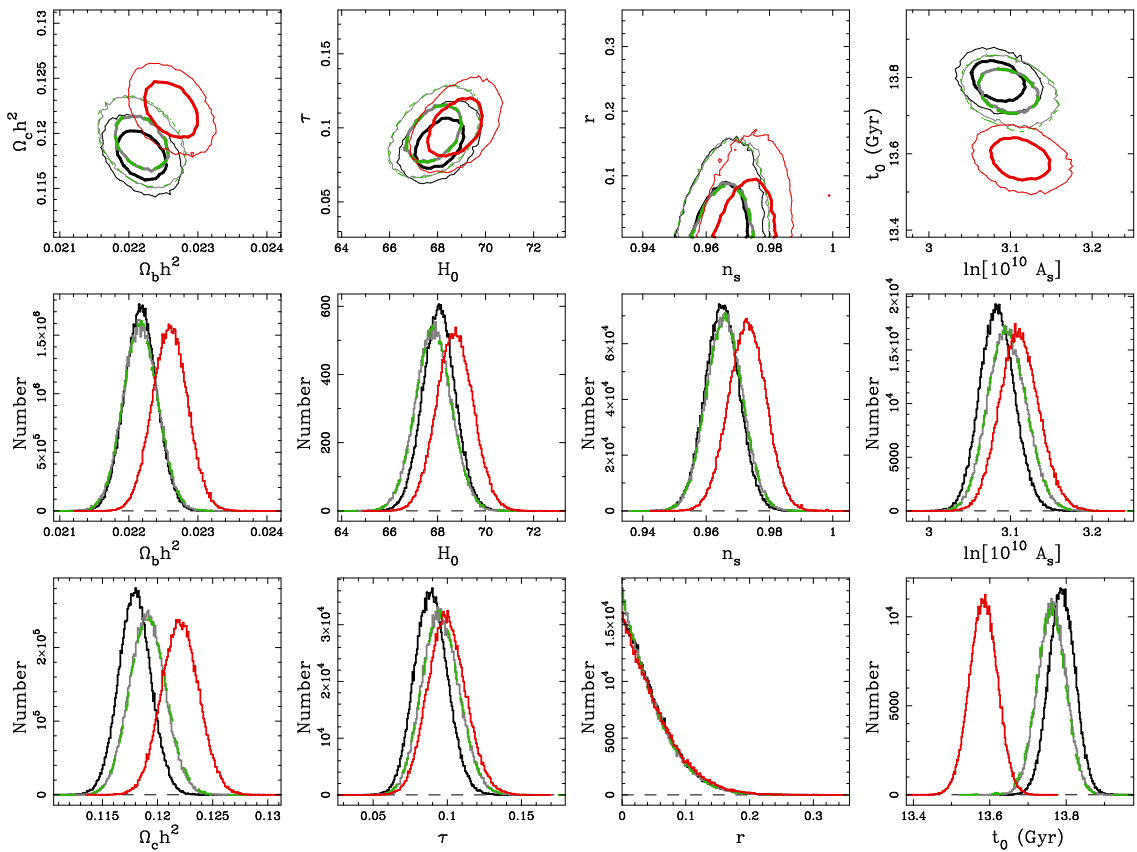,width=150mm,clip=}}
\caption{Top: Two-dimensional likelihood contours favored by the recent
         observations (Planck+WP+BAO) for EDE-4 (with $\phi_0$,
         $A$, $\lambda$ as free parameters; green dashed) and EDE-5 ($\phi_0 = -5$, $A=0.01$,
         $\lambda=20$; red contours) models.
         The 68.3\% and 95.4\% confidence limits are indicated by the thick
         and thin solid curves, respectively.
         The results for the SDE (gray) and $\Lambda\textrm{CDM}$ (black
         contours) models are shown for comparison.
         Middle and bottom: Marginalized one-dimensional likelihood
         distributions for each cosmological parameter.
         }
\label{fig:cont_EDE_nt_all_planck}
\end{figure*}



\begin{table*}
\caption{Mean and standard deviation ($68.3$\% confidence limit) in the
         marginalized one-dimensional likelihood distribution for best-fit
         $\Lambda\textrm{CDM}$, SDE, EDE models constrained with the recent
         observational data sets (Planck+WP+BAO). For SDE model, $\lambda$
         is a free parameter. For EDE-4 model, free potential parameters are
         $\lambda$, $\phi_0$, and $A$.
         For tensor-to-scalar ratio $r$, the upper limit is presented.
}
\begin{ruledtabular}
\begin{tabular}{lcccc}
  Parameter & $\Lambda\textrm{CDM}$        &  SDE $(\phi_0=-3,A=100)$  & EDE-4                & EDE-5 $(\lambda=20,\phi_0=-1.5,A=0.01)$  \\[1mm]
\hline \\[-2mm]
  $100\Omega_{b}h^2$ & $2.219 \pm 0.023$   & $2.218 \pm 0.025$         & $2.218\pm 0.025$     & $2.261 \pm 0.025$ \\[+1mm]
  $\Omega_{c}h^2$    & $0.1180 \pm 0.0015$ & $0.1192 \pm 0.0016$       & $0.1192\pm 0.0017$   & $0.1222 \pm 0.0017$  \\[+1mm]
  $h$                & $0.6810 \pm 0.0068$ & $0.6786 \pm 0.0075$       & $0.6787\pm 0.0076$   & $0.6876 \pm 0.0076$ \\[+1mm]
  $\tau$             & $0.090 \pm 0.011$   & $0.096 \pm 0.013$         & $0.096\pm 0.013$     & $0.101 \pm 0.013$ \\[+1mm]
  $n_s$              & $0.9653 \pm 0.0054$ & $0.9660 \pm 0.0057$       & $0.9660\pm 0.0058$   & $0.9733 \pm 0.0059$ \\[+1mm]
  $r$                & $ < 0.054$          & $< 0.051$                 & $<0.054$             & $< 0.057$ \\[+1mm]
  $\ln[10^{10}A_s]$  & $3.085 \pm 0.021$   & $3.097 \pm 0.024$         & $3.098\pm 0.024$     & $3.112 \pm 0.024$ \\[+1mm]
  $t_0$ (Gyr)        & $13.790 \pm 0.035$  & $13.765 \pm 0.038$        & $13.761\pm 0.039$    & $13.585 \pm 0.036$ \\[+1mm]
\end{tabular}
\end{ruledtabular}
\label{tab:EDE_constraints_planck}
\end{table*}

\begin{figure*}
\mbox{\epsfig{file=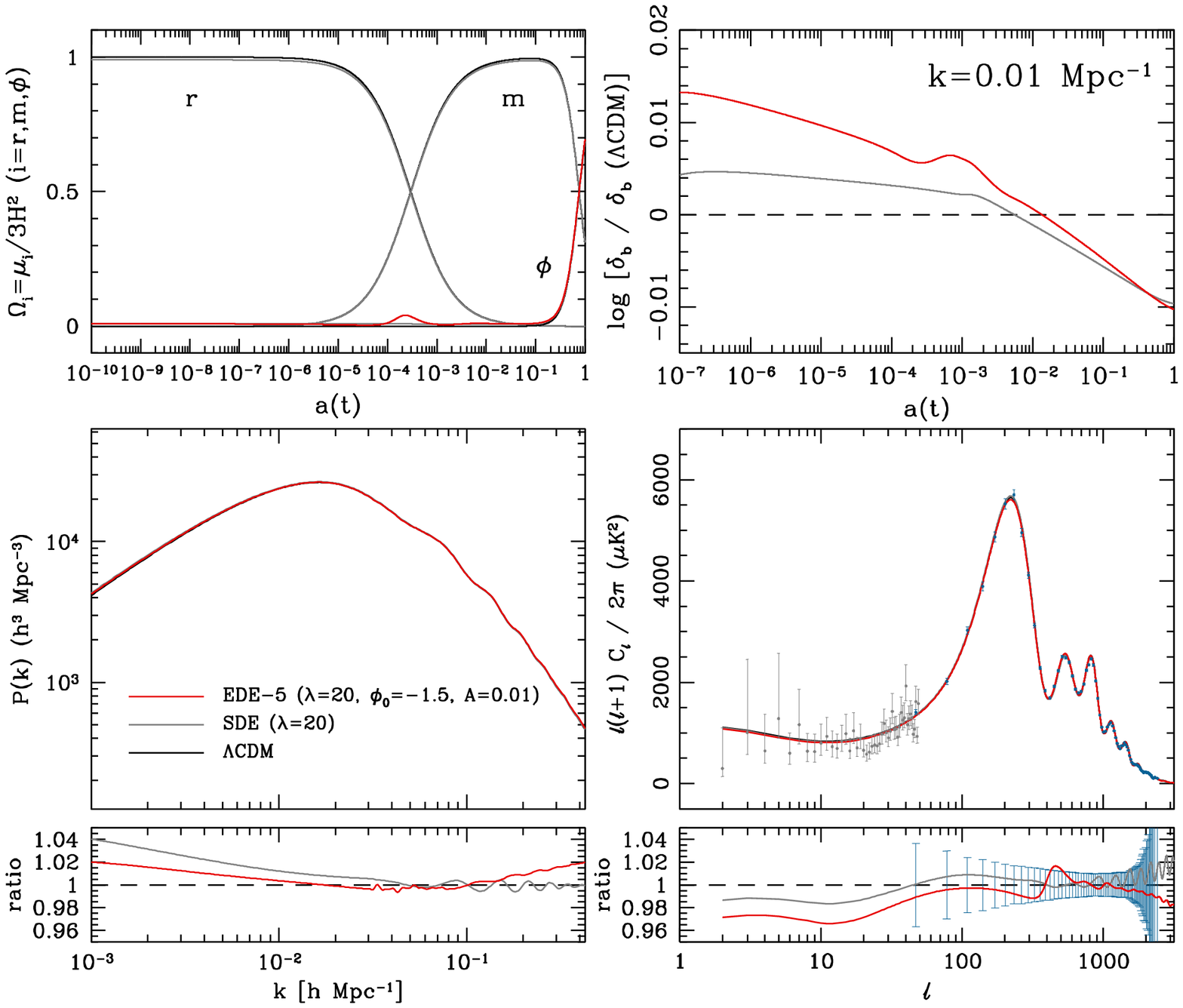,width=120mm,clip=}}
\caption{Evolution of background density parameters (top-left) and baryon
         density perturbations ($\delta_b \equiv \delta\mu_b /\mu_b$
         in the CDM-comoving gauge) at $k=0.01~\textrm{Mpc}^{-1}$
         relative to the $\Lambda\textrm{CDM}$ model (top-right),
         and baryonic matter and CMB anisotropy power spectra (bottom panels)
         for the EDE-5, SDE, and $\Lambda\textrm{CDM}$ models best-fitting
         Planck+WP+BAO data sets, with the same color codes as in
         Fig.\ \ref{fig:cont_EDE_nt_all_planck}.
         Note that the CMB temperature power spectrum is the sum
         of contributions from the scalar- and tensor-type perturbations.
         For CMB temperature power spectrum, the recent measurement
         from Planck data \cite{Planck-2013} has been added (gray and blue
         dots with error bars).
         In the ratio panels, we present the power ratio relative to
         the $\Lambda\textrm{CDM}$ prediction together with the fractional
         error bars of the observational data (only for Planck CMB angular
         power spectrum at small angular scales).
         }
\label{fig:bgpert_nt_bestfit_planck}
\end{figure*}

With the same numerical tools, we have explored the parameter constraints
using the recent CMB data from the Planck satellite \cite{Planck-2013}.
Due to the Planck's high precision up to small angular scales
($\ell \approx 3000$), even tighter constraints on model parameters are
expected. The Planck data includes the CMB temperature anisotropy angular
power spectrum, WMAP 9-year polarization data (WP) \cite{WMAP9},
and the cross-correlation between them \footnote{Based on observations
obtained with Planck (http://www.esa.int/Planck), an ESA science mission
with instruments and contributions directly funded by ESA Member States, NASA,
and Canada.}.
We use the Planck data (CAMspec version 6.2) and run the modified CosmoMC
software to obtain the likelihood distribution of cosmological parameters.
As in the Planck team's analysis, effects of unresolved foregrounds,
calibration, and beam uncertainties have been considered and the related 
parameters have been marginalized over \cite{Planck-2013}.
The pivot scale for the initial power spectrum amplitude has been set to
$k=0.05~\textrm{Mpc}^{-1}$.
As the large-scale structure data, we use the BAO measurements obtained 
from the Six-Degree-Field Galaxy Survey \cite{6dF}, the SDSS DR 7
\cite{SDSS-DR7}, and Baryon Oscillation Spectroscopic Survey DR 9
\cite{BOSS-DR9}. 

With the combined data sets (denoted as Planck+WP+BAO), we have constrained
the parameter space of spatially flat $\Lambda\textrm{CDM}$, SDE, and EDE
models that are favored by the observations.
The results are summarized in Fig.\ \ref{fig:cont_EDE_nt_all_planck} and
Table \ref{tab:EDE_constraints_planck}.

For SDE model, we have set $\lambda$ as a free parameter
to constrain the initial level of dark energy allowed at the early epoch.
The allowed range for this parameter is $\lambda > 17.9$, which corresponds
to the level of early dark energy $\Omega_{e} \lesssim 0.012$ ($95.4$\%
confidence limit). This result is similar to the recent constraint on the
fluid-based EDE density parameter obtained from the Planck observation
\cite{Planck-2013}. 
When Planck+WP+BAO data sets are used, the parameter constraints of SDE model
are very similar to those of $\Lambda\textrm{CDM}$ model.
The chi-square value at the best-fit position is
$\chi_{\textrm{min}}^2/2=4909.3$ for SDE and $4908.1$ for
$\Lambda\textrm{CDM}$ model.
The SDE model prefers slightly larger values of $\Omega_c h^2$, $\tau$, and
$\ln[10^{10} A_s]$ and smaller value of $h$ and $t_0$ (age) than
$\Lambda\textrm{CDM}$ model. The estimated parameters of both models are
consistent within $1 \sigma$ uncertainty, with small differences less than
one percent.

For EDE models, we have considered two cases. In the first case, the
observational data sets have been compared with an EDE model where the
potential parameters $\lambda$, $A$, $\phi_0$ are all freely varied together
with the conventional cosmological parameters, but with flat priors,
$10 \le\lambda\le 25$, $-2.3 \le \log_{10} A \le 1$, and
$-5 \le \phi_0 \le -1$ (hereafter EDE-4).
With the Planck+WP+BAO data, the potential parameters in the direction to
the upper (lower) bound of $\lambda$ and $A$ ($\phi_0$) are preferred
\footnote{Within the flat priors, the allowed ranges of potential parameters
are $\lambda > 18.2$, $\phi_0 < -1.62$, and $\log_{10} A > -2.11$
(95.4\% confidence limits).},
and the resulting constraints on the conventional cosmological parameters are
essentially the same as those in the SDE case, with
$\chi_{\textrm{min}}^2/2=4909.4$ (see Table \ref{tab:EDE_constraints_planck}).
The parameter constraint results above suggest that the Planck+WP+BAO data
sets strongly favor the $\Lambda\textrm{CDM}$ model than the scalar-field
based EDE model. 

Although the Planck data provides a really tight constraint on the
$\Lambda\textrm{CDM}$ model, as a second case we consider a particular EDE
model in which the early episodic domination of dark energy occurs near the
radiation-matter equality (hereafter EDE-5).
We have designed this model by setting $A=0.01$,
$\phi_0=-1.5$ for $\lambda=20$ to make the dark energy density have the
maximum strength $\Omega_{\phi}=0.04$ around the radiation-matter equality.
The choice of $\lambda=20$ is to fix the level of early dark energy,
$\Omega_{\phi, i}=0.01$, based on the fluid-based EDE constraint from Planck
data \cite{Planck-2013}. The results are presented in Fig.\
\ref{fig:cont_EDE_nt_all_planck} and Table \ref{tab:EDE_constraints_planck}.

Figure \ref{fig:bgpert_nt_bestfit_planck} shows the evolution of background
and perturbation quantities, and the matter and CMB power spectra of the
best-fit $\Lambda\textrm{CDM}$, SDE (with $\lambda=20$ fixed), and EDE-5
($\phi_0 =-1.5$, $A=0.01$, $\lambda=20$) models. 
The best-fit designed EDE model has a small bump in the dark energy density
parameter with the maximum at $a=2.3\times 10^{-4}$ or redshift $z=4350$
(top-left panel). The evolution of baryon density perturbation behaves in
a similar way to that of EDE-2 model, but now with the smaller difference
from $\Lambda\textrm{CDM}$ model.
The designed EDE model gives a poorer fit, with the minimum chi-square value
($\chi_{\textrm{min}}^2/2=4911.1$) larger than $\Lambda\textrm{CDM}$ model.
However, in the matter and CMB power spectra, the deviation of the best-fit
EDE model from $\Lambda\textrm{CDM}$ one is small, showing only a few \%
difference at all scales (see Fig.\ \ref{fig:bgpert_nt_bestfit_planck}
bottom panels showing the power spectrum ratio relative to
$\Lambda\textrm{CDM}$ model).
The EDE model predictions are consistent with observational data within
uncertainties, except for the slightly larger amplitude of CMB temperature
power spectrum around the second acoustic peak, which is the primary reason
for the deviation from $\Lambda\textrm{CDM}$ model.

The deviation of parameter constraints between this model (EDE-5) and
$\Lambda\textrm{CDM}$ is still quite small as expected, but with noticeable
differences. We note the similar behavior of parameter deviation as seen
in the case of EDE-2 model (constrained with WMAP7+LRG+$H_0$ data). 
For example, the EDE-5 model favors the baryon (CDM) density that is 1.9\%
(3.6\%) higher than the $\Lambda\textrm{CDM}$ best-fit value, with a
statistical deviation by $1.8\sigma$ ($2.8\sigma$).
Besides, the derived cosmic age ($t_0=13.585\pm 0.036~\textrm{Gyr}$)
is smaller than the $\Lambda\textrm{CDM}$ best-fit value
($t_0=13.790\pm 0.035~\textrm{Gyr}$) with $1.5$\% difference (deviating
from $\Lambda\textrm{CDM}$ model by $5.8\sigma$).
Such deviations mentioned above become smaller as we choose the weaker
episodic domination of dark energy in our EDE model.

Considering all the cases of $\Lambda\textrm{CDM}$, SDE and EDE models
constrained with the Planck data, we found that the Planck data strongly
favors the $\Lambda\textrm{CDM}$ model and only a limited amount of dark
energy with episodic nature is allowed.
Although the statistical deviation from $\Lambda\textrm{CDM}$ model is small,
our results still imply that a different parameter estimation with some
deviations from the $\Lambda\textrm{CDM}$ model can be obtained based on
the same observational data by introducing the ephemerally dominating dark
energy at early epoch.

\section{Summary and Conclusion}
\label{sec:conclusions}

In this paper, we investigate the observational effects of early
episodically dominating dark energy based on a minimally coupled
scalar field with the Albrecht-Skordis potential.
Our results show that the episodic domination of the dark energy component
can affect the cosmological parameter constraints significantly, compared
with the conventional estimation based on the $\Lambda$CDM model.
For the WMAP data, we found that the EDE dominating after the radiation era
(EDE-2, EDE-3) affects the growth of density perturbations
(Fig.\ \ref{fig:bgpert_nt_bestfit}), consequently modifying the observationally
favored parameter space, compared with the $\Lambda\textrm{CDM}$ model
(Fig.\ \ref{fig:cont_EDE_nt_all});
one can make a model that prefers the shorter cosmic age (EDE-2)
or the existence of tensor-type perturbation (EDE-3).

For the Planck data, the effect of early dark energy with episodic nature
should be sufficiently suppressed to be consistent with observational data (EDE-4).
However, we note the similar trends as seen in the EDE models constrained
with WMAP data (Fig.\ \ref{fig:cont_EDE_nt_all_planck}).
In the presence of transiently dominating dark energy at the early epoch,
the estimated cosmological parameters can deviate from the
currently known $\Lambda\textrm{CDM}$-based parameters with the percent level
difference (EDE-5; Fig.\ \ref{fig:bgpert_nt_bestfit_planck}).

We note that this interesting phenomenon is not seen in the case of 
the conventional dark energy model where the estimated cosmological
parameters are very similar to $\Lambda\textrm{CDM}$ parameters
\cite{Komatsu-etal-2011}.
Our model can be considered as an example where alternative cosmological
parameter estimations are allowed based on the same observations
even in Einstein's gravity.

\section*{Appendix: Initial conditions for background evolution of EDE models}

Here we present initial conditions for the background evolution of
several EDE models considered in this paper. It should be noted that all the scalar field
parameters and variables are expressed under the unit where $c \equiv 1$ and $8\pi G \equiv 1$.
Besides, in the numerical calculation, the potential parameters $V_0$ and $V_1$
in Eq.\ (\ref{eq:V}) are expressed in unit of $H_0^2$,
that is, $\hat{V}_0=V_0/H_0^2$ and $\hat{V}_1=V_1/H_0^2$. We have set $\hat{V}_0\equiv 1$ and $\beta=0$.
Table \ref{tab:EDE_IC} lists scaling initial conditions of background scalar field variables
($\phi_i$, $\phi_i^\prime$) at the initial epoch $a_i=10^{-10}$, together with potential
parameters of EDE models considered in Figs.\ \ref{fig:bg_example}, \ref{fig:bgpert_examples},
\ref{fig:bgpert_nt_bestfit}, and \ref{fig:bgpert_nt_bestfit_planck}.
In Fig.\ \ref{fig:bg_example}, we assume the conventional cosmological parameters as those of
spatially flat $\Lambda\textrm{CDM}$ model from the WMAP 7-year result (\cite{Larson-etal-2011};
$\Omega_b h^2=0.02260$, $\Omega_c h^2=0.1123$, $h=0.704$, $T_0=2.725~\textrm{K}$, $N_\nu = 3.04$
[number of species of massless neutrinos]),
while in Fig.\ \ref{fig:bgpert_examples}, as those from the Planck result (\cite{Planck-2013};
$\Omega_b h^2 = 0.022161$, $\Omega_c h^2=0.11889$, $h=0.6777$, $T_0=2.7255~\textrm{K}$,
$N_\nu=3.046$).

\begin{table*}
\caption{Scaling initial conditions of scalar field variables and potential parameters of EDE models.}
\begin{ruledtabular}
\begin{tabular}{llcccccc}
  Figure &  Model   &$\lambda$ & $\phi_0$ & $A$       & $\hat{V}_1$ & $\phi_i$     & $\phi_i^{\prime}$   \\[1mm]
\hline \\[-2mm]
  Fig.\ \ref{fig:bg_example}
          &  SDE    & $10$     & $-4$     & $100$     &  $2.14417$  &  $-7.48116$  & $0.397531$   \\[+1mm]
          &  EDE    & $10$     & $-4$     & $0.01$    &  $2.14134$  &  $-7.68262$  & $0.379410$   \\[+1mm]
\hline \\[-2mm]
  Fig.\ \ref{fig:bgpert_examples}
          &  SDE    & $10$     & $-3$     & $100$     &  $2.03666$  &  $-7.48180$  & $0.397036$   \\[+1mm]
          &  EDE    & $10$     & $-5$     & $0.0187$  &  $2.03398$  &  $-7.74537$  & $0.372901$   \\[+1mm]
          &  EDE    & $10$     & $-4$     & $0.0183$  &  $2.03309$  &  $-7.68996$  & $0.379460$   \\[+1mm]
          &  EDE    & $10$     & $-3$     & $0.0162$  &  $2.03186$  &  $-7.64613$  & $0.383504$   \\[+1mm]
          &  EDE    & $10$     & $-2$     & $0.0148$  &  $2.03253$  &  $-7.60993$  & $0.386237$   \\[+1mm]
\hline \\[-2mm]
  Fig.\ \ref{fig:bgpert_nt_bestfit} 
          &  SDE    & $10$     & $-3$     & $100$     &  $2.13829$  &  $-7.47420$  & $0.397040$   \\[+1mm]
          &  EDE-1  & $10$     & $-4$     & $0.01$    &  $2.10625$  &  $-7.68552$  & $0.379425$   \\[+1mm]
          &  EDE-2  & $10$     & $-3$     & $0.02$    &  $2.07960$  &  $-7.62556$  & $0.383436$   \\[+1mm]
          &  EDE-3  & $10$     & $-2$     & $0.04$    &  $2.15929$  &  $-7.60357$  & $0.386232$   \\[+1mm]
\hline \\[-2mm]
  Fig.\ \ref{fig:bgpert_nt_bestfit_planck} 
          &  SDE    & $20$     & $-3$     & $100$     &  $2.05858$  &  $-3.67975$  & $0.199865$   \\[+1mm]
          &  EDE-5  & $20$     & $-1.5$   & $0.01$    &  $2.07378$  &  $-3.82015$  & $0.191751$   \\[+1mm]
\end{tabular}
\end{ruledtabular}
\label{tab:EDE_IC}
\end{table*}

%
%
\acknowledgements

C.G.P. was supported by Basic Science Research Program through the National
Research Foundation of Korea (NRF) funded by the Ministry of Science, ICT
and Future Planning (No.\ 2013R1A1A1011107) and was partly supported
by research funds of Chonbuk National University in 2012.
J.H.\ was supported by Basic Science Research Program through the NRF of Korea
funded by the Ministry of Science, ICT and future Planning
(No.\ 2013R1A1A2058205).
H.N.\ was supported by NRF of Korea funded by the Korean Government
(No.\ 2012R1A1A2038497).

\def\and{{and }}


\end{document}